\documentclass[a4paper,11pt]{article}

\usepackage[left=2.5cm,right=2.5cm,top=2.5cm,bottom=2.5cm]{geometry}
\usepackage{graphicx,amssymb,amsmath,amsthm,mathrsfs,setspace,subcaption,cite,authblk,float}

\usepackage[colorlinks=true,bookmarks=true]{hyperref}
\usepackage{breakurl} %%This package must be after hyperref

\usepackage{mathtools}

\DeclarePairedDelimiterX\braket[2]{\langle}{\rangle}{#1 \delimsize\vert #2}

\hypersetup{citecolor=blue}
\newcommand{\dif}{\mathrm{d}}
\newcommand{\Eqref}[1]{(\ref{#1})}
\newcommand{\half}{\frac{1}{2}}
\newcommand{\expo}[1]{\mathrm{e}^{#1}}
\newcommand{\brac}[1]{\left(#1 \right)}
\newcommand{\sbrac}[1]{\left[#1\right]}

\numberwithin{equation}{section}
 
\begin{document}

\title{Model for an electrostatic capacitor in Einstein--Maxwell theory}

\author{Ian Khai-Shuen Ng\footnote{Email: phy2009960@xmu.edu.my}}

\author{Wei Zheng Choo\footnote{Email: eee2004046@xmu.edu.my}}

\author{Yen-Kheng Lim\footnote{Email: yenkheng.lim@gmail.com, yenkheng.lim@xmu.edu.my}}

\affil{\normalsize{\textit{Department of Physics, Xiamen University Malaysia, 43900 Sepang, Malaysia}}}

\date{\normalsize{\today}}

\renewcommand\Authands{, and }

\maketitle 
% \begin{center}
%  \Large{\textbf{Model for an electrostatic capacitor in Einstein--Maxwell theory}}
% \end{center}

\begin{abstract}
 A general relativistic model of a parallel-plate electrostatic capacitor is presented. The spacetime is a solution to the Einstein--Maxwell equations and involves class of solution previously studied by Vesel\'{y} and \v{Z}ofka (V\v{Z}). In particular, the parts containing curvature singularities are cut out and the remaining regular section is glued to asymptotically-Minkowski spacetimes. In essence, this results in a curved electro-vacuum V\v{Z} spacetime sandwiched on both sides by exterior spacetimes with vanishing electromagnetic fields. Junction conditions require the presence of charged matter on the boundaries. We interpret this configuration as a parallel-plate capacitor with gravitational effects induced by the strong electric fields. The spherical capacitor is briefly considered.
\end{abstract}

\section{Introduction} \label{sec_intro}

The capacitor is one of the simplest electrical device that is well described by Gauss' law in elementary electrostatics. However, a sufficiently strong electric field also works as a source of gravity, curving the spacetime in its presence. As such, the main subject of this paper is to consider a parallel-plate capacitor which takes the curvature of spacetime into account, using Einstein--Maxwell gravity. Previously, other gravitational counterparts of electromagnetic systems have been explored under different contexts. For instance, the \emph{cosmic solenoid} was studied by Davidson and Karasik in \cite{Davidson:1999fa}, and the \emph{rechargeable black-hole battery} by Mai and Yang in \cite{Mai:2022pgf}.

In this paper, the `cosmic capacitor' is obtained from a slight modification of a particular electro-vacuum solution to Einstein--Maxwell equations. The magnetic version of this solution was one of the many cylindrically-symmetric spacetimes obtained in the review \cite{Bronnikov:2019clf}. Subsequently it was further studied and generalised to include a cosmological constant by Vesel\'{y} and \v{Z}ofka (V\v{Z}) \cite{Vesely:2021jlc}. See also \cite{Zofka:2019yfa,Vesely:2019ajp} for related developments.

Further properties of this spacetime (such as its causal structure, particle motion, and thin shell sources, among others) was studied in further detail in Sec.~2.4 of \cite{Vesely:2022vws}. Here, let us only review the features that are relevent for our present purposes. Specifically, we are interested in the electric version of this solution with zero cosmological constant, for which the metric and gauge potential is\footnote{This solution is found in Eq.~(3.16) of Ref.~\cite{Bronnikov:2019clf}, Eq.~(13) of \cite{Vesely:2021jlc}, or Eq.~(2.176) of \cite{Vesely:2022vws}.}
\begin{subequations}\label{VZspacetime}
\begin{align}
 \dif s^2&=-\frac{\dif t^2}{1-\sigma^2x^2}+\dif x^2+\brac{1-\sigma^2x^2}\brac{\expo{2\arcsin(\sigma x)}\dif z^2+\expo{-2\arcsin(\sigma x)}\dif w^2},\\
  A&=-\frac{\sigma x}{\sqrt{1-\sigma^2x^2}}\,\dif t.
\end{align}
\end{subequations}
The Faraday tensor is obtained by $F=\dif A$. The domain for the coordinates are $t,z\in\mathbb{R}$, $x\in\brac{-1/\sigma,1/\sigma}$, and $w\in[0,2\pi)$. There are curvature singularities at $x\rightarrow\pm1/\sigma$ \cite{Vesely:2021jlc}. The flat spacetime limit is $\sigma\rightarrow0$, for which $A=0$ and the metric becomes
\begin{align*}
 \dif s^2&=-\dif t^2+\dif x^2+\dif z^2+\dif w^2.
\end{align*}
Note that since $w$ is a periodic coordinate, this flat limit is a three-dimensional Minkowski spacetime times a circle, $\mathbb{R}^{2,1}\times S^1$.

If we instead consider the solution locally identical to \Eqref{VZspacetime}, but with $w\in\mathbb{R}$ taken to be  a non-periodic coordinate, the global nature of the solution is different. For this case, the flat limit $\sigma\rightarrow0$ is four-dimensional Minkowski, $\mathbb{R}^{3,1}$, with no compact dimensions.

% with a trivial renaming $r\rightarrow x$ and $\phi\rightarrow w$. However, the global nature of the solution is different if $w$ takes the domain $w\in\mathbb{R}$. The other coordinates take the same domains as \Eqref{VZspacetime}, namely $t,z\in\mathbb{R}$ and $x\in\brac{-1/\sigma,1/\sigma}$. For this case, the flat limit $\sigma\rightarrow0$ is four-dimensional Minkowski, $\mathbb{R}^{3,1}$, with no compact dimensions.

In Ref.~\cite{Vesely:2022vws}, Vesel\'{y} considered a shell source which acts as a boundary that separates the V\v{Z} spacetime with another instance of itself. In this paper, we will also consider shell sources, but by sandwiching the V\v{Z} spacetime between two exterior, asymptotically-flat spacetimes. More specifically, we consider \Eqref{VZspacetime} with $w\in\mathbb{R}$ and remove the curvature singularities by cutting the spacetime at $x=\pm a$, where $a<\frac{1}{\sigma}$. Then each side is glued with asymptotically-Minkowski spacetimes with vanishing electromagnetic fields. The junction conditions \cite{PoissonToolkit,Israel:1966rt} then requires the presence of oppositely-charged matter at the surfaces $x=\pm a$. We interpret this configuration as a model of an electrostatic parallel-plate capacitor in Einstein--Maxwell theory. Indeed, in the limit of weak gravity, the solution reduces to that of a uniform electric field between two oppositely charged planes. We note in passing that charged shell methods have been applied in other contexts, such as in Refs.~\cite{Gurses:1998xa,Gurses:2005ns,Garcia:2011aa,Culetu:2015oha}.

The rest of this paper is organised as follows. In Sec.~\ref{sec_eom}, we describe the spacetime metrics and gauge potentials. In the same section we also investigate the properties of the surface stress tensor as required by the junction conditions. Physical orders of magnitudes, including the `capacitance' of the system, are calculated in Sec.~\ref{sec_capacitance}. The behaviour of charged test particles are investigated in Sec.~\ref{sec_particles}. In Sec.~\ref{sec_spherical} we briefly consider the spherical capacitor. The paper concludes in Sec.~\ref{sec_conclusion}. We will mainly work in geometric units where $G=c=1$, except in Sec.~\ref{sec_capacitance} where we briefly convert some physical quantities to SI units. We follow the conventions of Poisson's book \cite{PoissonToolkit} for curvature and stress tensors.

\section{Bulk solutions and junction conditions} \label{sec_eom}

\begin{figure}
 \centering 
 \includegraphics[scale=0.6]{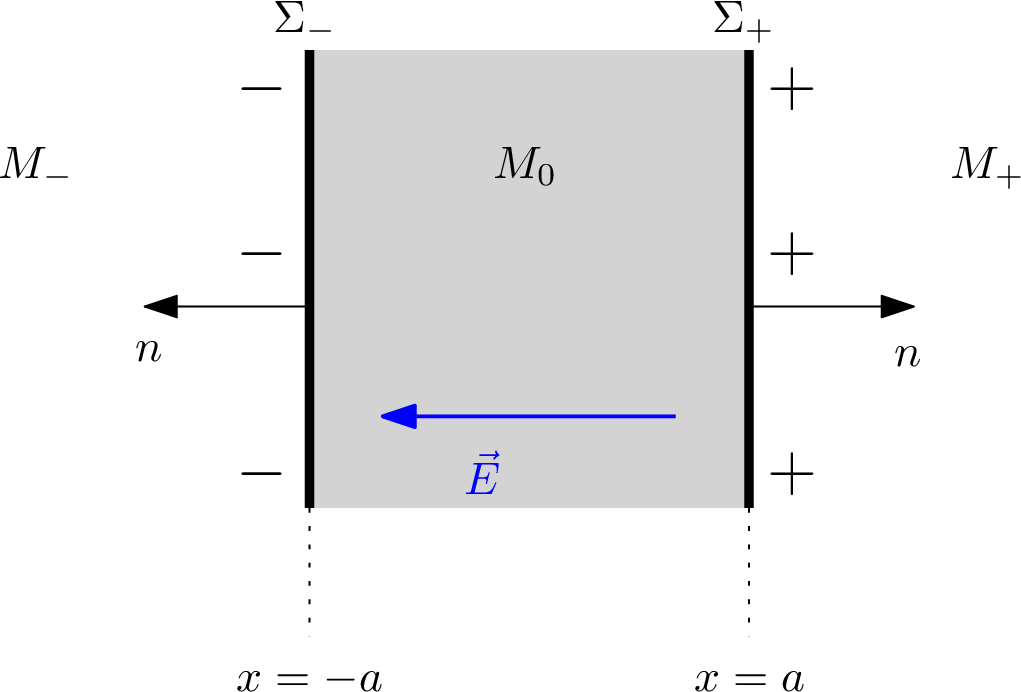}
 \caption{Sketch of the spacetime depicting a cosmic capacitor. The shaded region represents the domain $M_0$ where an electric field points from $\Sigma_+$ to $\Sigma_-$. The domains $M_\pm$ are asymptotically-Minkowski spacetimes. The surfaces $\Sigma_\pm$ carries positive and negative charges, respectively.}
 \label{fig_capacitor}
\end{figure}

Consider a four-dimensional spacetime $M$ partitioned into three parts, $M=M_-\cup M_0\cup M_+$. Let $\Sigma_-$ be the common boundary between $M_-$ and $M_0$. Similarly let $\Sigma_+$ be the common boundary between $M_0$ and $M_+$, as shown in Fig.~\ref{fig_capacitor}. Both surfaces $\Sigma_\pm$ are taken to be time-like. Our convention for the intrinsic and extrinsic geometry is as follows: We will let $x^\mu$  denote bulk coordinates on $M_i$, where $i$ stands for $i\in\{-,0,+\}$, and $y^a$ be coordinates on the surfaces $\partial M_i$. The induced metric on $\partial M_i$ is $h_{ab}=g_{\mu\nu}e^\mu_ae^\nu_b$, where $e^\mu_a=\frac{\partial x^\mu}{\partial y^a}$ and $g_{\mu\nu}$ is the bulk metric. If $n^\mu$ is the outward-pointing unit normal at a boundary $\partial M_i$, the extrinsic curvature is $K_{ab}=e^\mu_a e^\nu_b\nabla_\mu n_\nu$. The trace of the extrinsic curvature is $K=K_{ab}h^{ab}$.

The action for the present model is 
\begin{align}
 I=I_{M_-}+I_{M_0}+I_{M_+}+I_{\Sigma}+I_{S}, \label{action}
\end{align}
where
\begin{align}
 I_{M_i}&=\frac{1}{16\pi}\int_{M_i}\dif^4x\sqrt{-\det g}\brac{R-F^2}+\int_{M_i}\dif^4x\sqrt{-\det g}\;\mathcal{L}_i,\quad i=-,0,+,
\end{align}
where $\mathcal{L}_i$ is the Lagrangian for the corresponding matter source in the bulk $M_i$. 
Here $R$ is the Ricci scalar and we have used the notation $F^2=F_{\mu\nu}F^{\mu\nu}$, where $F_{\mu\nu}=\nabla_\mu A_\nu-\nabla_\nu A_\mu$ are the components of the Faraday tensor $F=\dif A$, with $A=A_\mu\,\dif x^\mu$ being the electromagnetic four-potential. The boundary terms are 
\begin{align}
 I_{\Sigma}=\frac{1}{8\pi}\int_{\Sigma_-\cup\Sigma_+}\dif^3y\sqrt{-\det h}\;K,
\end{align}
where $K$ is the extrinsic curvature and $\det h$ is the determinant of the metric induced on $\Sigma_\pm$. The term $I_S$ is the action describing the `capacitor plates',
\begin{align}
 I_S&=\int_{\Sigma_-\cup\Sigma_+}\dif^3y\sqrt{-\det h}\;\mathcal{L}_S,
\end{align}
where $\sqrt{-\det h}\;\mathcal{L}_S$ represents the Lagrangian density for the matter source.

The equations of motion in the bulk (that is, away from $\Sigma_\pm$) are obtained by varying \Eqref{action} with respect to $g^{\mu\nu}$ and $A_\mu$, giving the Einstein--Maxwell equations in the bulk $M_\pm$, $M_0$.
\begin{subequations}\label{eom_bulk}
\begin{align}
 R_{\mu\nu}-\half R g_{\mu\nu}&=2F_{\mu\lambda}{F_\nu}^\lambda-\half F^2g_{\mu\nu}+8\pi T_{\mu\nu},\\
 \nabla_\mu F^{\mu\nu}&=0.
\end{align}
\end{subequations}
where $T_{\mu\nu}=-2\frac{\delta\mathcal{L}_i}{\delta g^{\mu\nu}}+\mathcal{L}_i g_{\mu\nu}$ is the stress-energy tensor on $M_i$. On the surfaces $\Sigma_\pm$, we vary \Eqref{action} with respect to $h^{ab}$ and $A_a=e^\mu_a A_\mu$, the latter of which is the electromagnetic four-potential projected onto the surfaces. The resulting equations of motion are
\begin{subequations}
\begin{align}
 -\frac{1}{8\pi}\brac{[K_{ab}]-[K]h_{ab}}&=\mathcal{T}_{ab},\label{surf_ST}\\
 \frac{1}{4\pi}[F_{\mu\nu}]n^\mu e^\nu_bh^{ba}&=\mathcal{J}^a,\label{surf_charge}
\end{align}
\end{subequations}
where the surface stress tensor and surface current are, respectively,
\begin{align}
 \mathcal{T}_{ab}&=-2\frac{\delta\mathcal{L}_S}{\delta h^{ab}}+\mathcal{L}_Sh_{ab},\quad \mathcal{J}^a=-\frac{\delta\mathcal{L}_S}{\delta A_a}.
\end{align}
Here we have used the notation $[X]=X(M_\pm)|_{\Sigma_\pm}-X(M_0)|_{\Sigma_\pm}$ for the jump of a tensor quantity $X$ across $\Sigma_\pm$. 

We now specify the explicit solution which describes the capacitor. First, the solution on $M_0$, the region between the `plates' is taken to be the electric version of Vesel\'y and \v{Z}ofka's solution \cite{Vesely:2021jlc}. For convenience, let us write it in the form
% \begin{subequations} \label{soln_M0}
% \begin{align}
%  \dif s_0^2&=-\frac{\dif t^2}{1-\sigma^2x^2}+\dif x^2+\brac{1-\sigma^2x^2}\brac{\expo{2\arcsin(\sigma x)}\dif z^2+\expo{-2\arcsin(\sigma x)}\dif w^2},\label{metric_M0}\\
%   A&=-\frac{\sigma x}{\sqrt{1-\sigma^2x^2}}\,\dif t,
% \end{align}
% \end{subequations}
\begin{subequations} \label{soln_M0}
\begin{align}
 \dif s_0^2&=-\expo{-2V}\dif t^2+\dif x^2+\expo{2V}\brac{\expo{2U}\dif z^2+\expo{-2U}\dif w^2},\label{metric_M0}\\
  A&=-\sigma x\expo{-V}\,\dif t,
\end{align}
\end{subequations}
where
\begin{align}
 V=\half\ln\brac{1-\sigma^2x^2},\quad U=\arcsin(\sigma x).
\end{align}
This solves the bulk Einstein--Maxwell equations \Eqref{eom_bulk} for $T_{\mu\nu}=0$. The domains of the coordinates are taken to be $t,z,w\in\mathbb{R}$ and $x\in(-a,a)$ for some positive $a$. There are no curvature singularities as long as $a<\frac{1}{|\sigma|}$. The boundaries $\Sigma_\pm$ are approached as $x\rightarrow\pm a$.

For regions `outside' the plates, let us take a metric ansatz of the form 
\begin{subequations}
\begin{align}
 \dif s_\pm^2&=-\expo{-2V_\pm}f_\pm(x)\dif t^2+\expo{2V_\pm}f_\pm(x)^{-1}\dif x^2+\expo{2V_\pm+2U_\pm}h_\pm(x)\brac{\dif z^2+\expo{-4U_\pm}\dif w^2},\\
     A&=\chi_\pm\expo{-V_\pm}\dif t,
\end{align}
\end{subequations}
where 
\begin{align*}
 V_\pm&\equiv V(x=\pm a)=\half\ln\brac{1-\sigma^2a^2},\quad U_\pm\equiv U(x=\pm a)=\pm\arcsin(\sigma a)
\end{align*}
are constants and $f_\pm(x)$ and $h_\pm(x)$ are functions of $x$. The domain of $x$ is $x<-a$ for $M_-$ and $x> a$ for $M_+$. We also take the electric field to vanish outside the plates so $\chi_\pm$ are constants. By the Einstein equations, the stress tensor components on $M_\pm$ are respectively
\begin{align}
 T^\pm_{tt}=\bar{\rho}^\pm\expo{-2V_\pm}f_\pm,\quad T^\pm_{xx}=p^\pm_x\expo{2V_\pm}f_\pm^{-1},\quad T_{ww}^\pm=p^\pm_\perp\expo{2V_\pm+2U_\pm},\quad T_{zz}=p^\pm_\perp\expo{2V_\pm-2U_\pm},
\end{align}
with
\begin{subequations}
 \begin{align}
  \bar{\rho}^\pm&=\frac{1}{16\pi}\expo{-2V_\pm}\brac{-\frac{h_\pm''}{h_\pm}-\frac{f_\pm'h_\pm'}{h_\pm}+\frac{f_\pm h_\pm'^2}{h_\pm^2}},\\
  \bar{p}_x^\pm&=\frac{1}{16\pi}\expo{-2V\pm}\brac{\frac{f_\pm h_\pm'^2}{2h_\pm}+f_\pm'h_\pm'},\\
  \bar{p}_w^\pm&=\frac{1}{16\pi}\expo{-2V_\pm}\brac{f_\pm''+\frac{h_\pm''}{h_\pm}+\frac{f_\pm'h_\pm'}{h_\pm}-\frac{f_\pm h_\pm'^2}{h_\pm^2}}.
 \end{align}
\end{subequations}

This metric ansatz was so chosen such that the continuity of the metric across the boundaries $\Sigma_\pm$ is satisfied by the condition 
\begin{align}
 f_\pm(\pm a)=h_\pm(\pm a)=1. \label{continuity}
\end{align}
Let us further suppose that the exterior spacetimes are asymptotically flat, so that 
\begin{align}
 \lim_{x\rightarrow\pm\infty}f_\pm(x)=\mathrm{constant},\quad \lim_{x\rightarrow\pm\infty}h_\pm(x)=\mathrm{constant}. \label{asymptotically_flat}
\end{align}
The continuity of the electric potential across $\Sigma_\pm$ then fixes $\chi_\pm$ to be 
\begin{align*}
 \chi_\pm=\frac{\mp\expo{V\pm}\sigma a}{1-\sigma^2 a^2}.
\end{align*}
% For the regions `outside' the plates, we take the metrics on $M_\pm$ to be Minkowski spaces:
% \begin{align*}
%  \dif s^2_\pm&=-\dif t_\pm^2+\dif x^2+\dif z_\pm^2+\dif w_\pm^2,\\
%  A&=\chi_\pm\dif t_\pm,
% \end{align*}
% where $\chi_\pm$ are constants, hence the electric field vanishes. The domain of $x$ is $x<-a$ on $\dif s_-^2$ and $x>a$ on $\dif s_+^2$. To ensure the continuity of the metric across $\Sigma_\pm$, we scale the coordinates $(t_\pm,z_\pm,w_\pm)$ so that
% \begin{align*}
%  t_\pm=\expo{-V_\pm}t,\quad z_\pm=\expo{V_\pm+U_\pm}z,\quad w_\pm=\expo{V_\pm-U_\pm}w,
% \end{align*}

For concreteness, we shall view $\Sigma_\pm$ from the perspective of $M_0$, being the boundary with disconnected pieces $\partial M_0=\Sigma_+\cup\Sigma_-$. We take vector normal to the boundary, $n^\mu$, to be outward-pointing. Therefore, $n^\mu$ takes the explicit form
\begin{align}
 n^\mu=\left\{\begin{array}{cc}
               +\delta^\mu_x & \mbox{at $\Sigma_+$},\\
               -\delta^\mu_x & \mbox{at $\Sigma_-$}.
              \end{array}\right.
\end{align}
Let $y^a$ be the coordinates on $\partial M_0$ and $e^\mu_a=\frac{\partial x^\mu}{\partial y^a}$. The induced metric on the boundary $\partial M_0$ is
\begin{align}
 h_{ab}\dif y^a\dif y^b&=\left\{\begin{array}{cc}
                                 -\frac{\dif t^2}{1-\sigma^2a^2}+\brac{1-\sigma^2a^2}\brac{\expo{2\arcsin(\sigma a)}\dif z^2+\expo{-2\arcsin(\sigma a)}\dif w^2} & \mbox{ on }\Sigma_+,\\
                        -\frac{\dif t^2}{1-\sigma^2a^2}+\brac{1-\sigma^2a^2}\brac{\expo{-2\arcsin(\sigma a)}\dif z^2+\expo{2\arcsin(\sigma a)}\dif w^2} & \mbox{ on }\Sigma_-,
                                \end{array}\right.
\end{align}
with $g^{\mu\nu}=n^\mu n^\nu+h^{ab}e^\mu_a e^\nu_b$, where $h^{ab}$ is the inverse of the induced metric.
%The extrinsic curvature is $K_{ab}=e^\mu_a e^\nu_b\nabla_\mu n_\nu$, with its trace denoted by $K=h^{ab}K_{ab}$.

With these ingredients, we can now calculate the stress tensor on the surfaces $\Sigma_\pm$, which is given by $\mathcal{T}_{ab}=-\frac{1}{8\pi}\brac{[K_{ab}]-[K]h_{ab}}$. The resulting components are

\begin{align}
  \mathcal{T}^\pm_{tt}&=\rho^\pm\expo{-2V_\pm},\quad\mathcal{T}^\pm_{zz}=p_z^\pm\expo{2V_\pm\pm 2U_\pm},\quad \mathcal{T}^\pm_{ww}=p_w^\pm\expo{2V_\pm-2U_\pm}, 
\end{align}
where 
\begin{subequations}\label{ST_surface}
\begin{align}
 \rho^\pm&=\frac{1}{8\pi}\left.\brac{\pm 2V_\pm'\mp\expo{-V_\pm}h'_\pm}\right|_{\pm a},\\
 p_z^\pm&=\frac{1}{8\pi}\left.\brac{\pm U'_\pm\pm\half\expo{-V_\pm}\brac{h'_\pm+f'_\pm}}\right|_{\pm a},\\
 p_w^\pm&=\frac{1}{8\pi}\left.\brac{\mp U'_\pm\pm\half\expo{-V_\pm}\brac{h'_\pm+f'_\pm}}\right|_{\pm a}.
\end{align}
\end{subequations}
The notation $|_{\pm a}$ is to remind ourselves that on the surfaces $\Sigma_\pm$, the derivatives $f_\pm'$ and $h_\pm'$ are evaluated on $x=\pm a$, respectively. 

At this stage, our capacitor configuration has not been fully determined yet as the functions $f_\pm(x)$ and $h_\pm(x)$ are yet to be determined. So far, our requirements are the continuity across the boundaries (Eq.~\Eqref{continuity}) and an asymptotically flat exterior (Eq.~\Eqref{asymptotically_flat}). These fixes the values of $\{f_\pm(x),h_\pm(x)\}$ at $x=\pm a$ and $x\rightarrow\pm\infty$, respectively. 

The derivatives of these functions at $x=\pm a$ determines the surface stress tensors $\mathcal{T}^\pm_{ab}$, and the functions at the intermediate values of $x$ determines the bulk stress tensors $T^\pm_{\mu\nu}$. We shall demand that the metrics on $M_\pm$ is sourced by reasonable stress tensors both on the surfaces $\Sigma_\pm$ and on the bulk $M_\pm$. In particular, let us consider the \emph{null energy condition} (NEC) as a benchmark. Explicitly, we wish to check whether $\mathcal{T}^\pm_{ab}\ell^a\ell^b\geq0$ for any null vector $\ell^a$ on $\Sigma_\pm$, and $T_{\mu\nu}k^\mu k^\nu\geq0$ for any null vector $k^\mu$ on $M_\pm$.

On the surfaces $\Sigma_\pm$, the NEC is satisfied if 
\begin{align}
 \pm\left.\brac{f_\pm'-h_\pm'}\right|_{\pm a}\geq\lambda\geq0, \label{NEC1}
\end{align}
where $\lambda=\frac{4\sigma^2a+2\sigma\sqrt{1-\sigma^2a^2}}{\sqrt{1-\sigma^2a^2}}$. On the bulk $M_\pm$, it is satisfied if
\begin{align}
 f_\pm''-\frac{fh_\pm''}{h_\pm}&\geq0,\label{NEC2}\\
 \frac{h_\pm'^2}{h_\pm}-2h_\pm''&\geq0.\label{NEC3}
\end{align}
Thus a reasonable matter distribution can be found if functions $f_\pm(x)$ and $h_\pm(x)$ can be chosen to obey Eqs.~\Eqref{NEC1}--\Eqref{NEC3}. 

First, we note that if the exterior $M_\pm$ are pure Minkowski spacetimes, $f_\pm$ and $h_\pm$ are constants and \Eqref{NEC1} cannot be satisfied unless $\sigma=0$, which means the entire configuration $M_-\cup M_0\cup M_+$ is the trivial vacuum Minkowski spacetime. In other words, a capacitor with pure Minkowski exteriors necessarily requires exotic matter to construct. Thus we seek an appropriate non-zero exterior matter distribution which leads to non-constant $f_\pm$ and $h_\pm$.  By trial, an explicit configuration that satisfies all conditions are found as:
\begin{subequations}
\begin{align}
 f_\pm(x)&=1-k\tanh(\beta a)\pm k\tanh(\beta x),\\
 h_\pm(x)&=1+\expo{-\alpha a^2}\mp\frac{x}{a}\expo{-\alpha x^2}.
\end{align}
\end{subequations}
For this choice, the surface energy density $\rho^\pm$ is negative, so the \emph{weak} energy condition is not satisfied. However, the \emph{null} energy condition can be satisfied for the following choices of constants: First, the choice $\alpha=\frac{3}{2a^2}$ leads to \Eqref{NEC3} being satisfied, and $k=\frac{2\expo{-3/2}+\lambda a}{\beta a\brac{1-\tanh^2(\beta a)}}$ satisfies \Eqref{NEC1} by making it an equality. The remaining condition \Eqref{NEC2} can be satisfied by choosing an appropriately small $\beta$. For instance, $\beta=0.0001/a$ for $\sigma=0.8/a$. Recall that the domain for $x$ is $x>+a$ for the upper signs $f_+(x)$ and $h_+(x)$ for which they tend to constants as $x\rightarrow+\infty$, and $x<-a$ for the lower signs $f_-(x)$ and $h_-(x)$, where they also tend to constants as $x\rightarrow-\infty$. Therefore this configuration is asymptotically flat in both directions.

\section{Electrical quantities, orders of magnitude, and capacitance} \label{sec_capacitance}

In this section we will consider the properties of the electric field in the interior, $M_0$ along with the surface charges on $\Sigma_\pm$. The surface charge density can be found from Eq.~\Eqref{surf_charge}, which gives
\begin{align}
 \mathcal{J}^b=\pm\frac{1}{4\pi}\frac{\sigma}{\sqrt{1-\sigma^2a^2}}\delta^b_t \quad\mbox{ on }\Sigma_\pm.
\end{align}
We shall denote the time-like Killing vector of the spacetime by $\xi^\mu$. On $\Sigma_\pm$, it is correspondingly $\xi_a=e^\mu_a\xi_\mu$. Then, the charge contained in a spatial area $S_\pm\subset\Sigma_\pm$ normal to $\xi$ is
\begin{align}
  Q=-\int_{S_\pm}\dif^2x\sqrt{\gamma}\;\xi_a\mathcal{J}^a=\pm\int_{S_\pm}\dif w\dif z\;\sigma.
\end{align}
The charge per unit area on the surfaces $\Sigma_\pm$ is then
\begin{align}
 \Theta=\frac{Q}{\mathcal{A}}=\pm\sigma,
\end{align}
where $\mathcal{A}=\int_{S_\pm}\dif w\dif z$ is an arbitrary area on $\Sigma_\pm$.

The electric field vector is obtained from $E^i=-\xi^\mu {F_\mu}^i$. The only non-zero component is
\begin{align}
 E^x=-\frac{\sigma}{1-\sigma^2x^2}.
\end{align}

The electromagnetic potential one-form is $A=-\frac{\sigma x}{\sqrt{1-\sigma^2x^2}}\,\dif t$. To define the capacitance of the system, we wish to consider a coordinate-invariant quantity for the potential difference between the plates. To this end, consider a stationary time-like particle with four velocity $u^\mu=\brac{\sqrt{1-\sigma^2x^2},0,0,0}$. The potential energy of the particle in the field $F=\dif A$ is $U=-qu^\mu A_\mu=\pm q\sigma a$. We take the electric potential to be $V=U/q$. Therefore the potential difference between the two plates is
\begin{align*}
 \Delta V=2\sigma a.
\end{align*}

From the solution \Eqref{soln_M0}, we see that flat Minkowski spacetime (i.e., zero gravity limit) is recovered when the dimensionless parameter $\sigma a$ tends to zero. In the other direction, the metric deviates significantly from flat Minkowski spacetime if $\sigma a$ is sufficiently large. (i.e., close to $1$.) To get an idea of how strong the electric field needs to be for $\sigma a$ to be appreciably non-zero, we restore to SI units, which gives
\begin{align}
 \sigma=\brac{\frac{c^4}{4\pi G\epsilon_0}}^{-1/2}\tilde{\sigma},
\end{align}
where $\tilde{\sigma}$ is the electric field strength in standard units. In these units, the charge per area on the plates is 
\begin{align}
 \tilde{\Theta}=\brac{\frac{c^4\epsilon_0}{4\pi G}}^{1/2}\Theta=\brac{\frac{c^4\epsilon_0}{4\pi G}}^{1/2}\frac{\sigma}{\sqrt{1-\sigma^2a^2}}.
\end{align}
In SI units, the potential difference is 
\begin{align}
 \Delta\tilde{V}=\brac{\frac{c^4}{4\pi G\epsilon_0}}^{1/2}2a\tilde{\sigma}\sqrt{1-\sigma^2a^2}.
\end{align}

To get a rough estimate of physical quantities required to curve the spacetime with gravity, suppose we have the capacitor plates separated at $0.2\;\mathrm{m}=20\;\mathrm{cm}$. That is, $a=0.1\;\mathrm{m}$. The corresponding electric field, charge per area, and potential difference are
\begin{align*}
 \tilde{\sigma}&\sim 1.04\times 10^{28}\;\mathrm{Volt}\,\mathrm{m}^{-1},\\
 \tilde{\Theta}&\sim 2.92\times 10^{16}\;\mathrm{C}\,\mathrm{m}^{-2}, \\
 \Delta\tilde{V}&\sim 2.08\times 10^{26}\;\mathrm{Volt},
\end{align*}
showing that extremely strong electric fields and potentials are required to generate the V\v{Z} spacetime.

If we define the capacitance per area as the ratio of the charge per area with its potential difference,
\begin{align}
 \mathcal{C}&=\frac{\tilde{\Theta}}{\Delta\tilde{V}}=\epsilon_0\frac{\Theta}{\Delta V}=\frac{\epsilon_0}{2a},
\end{align}
which is the same expression as the standard non-relativistic capacitor.

\section{Behaviour of charged particles in the capacitor} \label{sec_particles}

The interplay between gravitational and electric fields can be revealed by studying the motion of charged particles between the plates. The motion of charged particles in the \emph{magnetic} V\v{Z} spacetime was studied in \cite{Vesely:2022vws}. In the electric solution depicting the capacitor interior, a charged particle experiences a Coulomb force that's always directed towards the plate of opposite charge, in addition to the gravitational field of the metric which gives a potential well.

To see this, let us consider a test particle of mass $m$ and charge $q$. Its motion is described by the Lagrangian $L=\frac{m}{2}g_{\mu\nu}\dot{x}^\mu\dot{x}^\nu+qA_\mu\dot{x}^\mu$. Explicitly,
\begin{align}
 L&=\frac{m}{2}\sbrac{-\frac{\dot{t}^2}{1-\sigma^2 x^2}+\dot{x}^2+\brac{1-\sigma^2 x^2}\brac{\expo{2\arcsin\sigma x}\dot{z}^2+\expo{-2\arcsin\sigma x}\dot{w}^2}}-\frac{q\sigma x\,\dot{t}}{\sqrt{1-\sigma^2x^2}}.
\end{align} 
Since $t$, $z$, and $w$ are cyclic variables of the Lagrangian, their canonical momenta $p_t=-E=\frac{\partial L}{\partial\dot{t}}$, $p_z=\frac{\partial{L}}{\partial\dot{z}}$, and $p_w=\frac{\partial L}{\partial\dot{w}}$ are constants, leading to the first order equations
\begin{subequations}\label{particle_eom}
\begin{align}
 \dot{t}&=\brac{1-\sigma^2x^2}\mathcal{E}-e\sigma\sqrt{1-\sigma^2x^2}\,x,\label{tdot}\\
 \dot{z}&=\expo{-2\arcsin\sigma x}\frac{P_z}{\brac{1-\sigma^2x^2}},\label{zdot}\\
 \dot{w}&=\expo{2\arcsin\sigma x}\frac{P_w}{\brac{1-\sigma^2x^2}},\label{wdot}
\end{align}
\end{subequations}
where we have introduced dimensionless variables for the energy, charge, and momenta:
\begin{align}
 \mathcal{E}=\frac{E}{m},\quad e=\frac{q}{m},\quad P_z=\frac{p_z}{m},\quad P_w=\frac{p_w}{m}.
\end{align}
The constraint $g_{\mu\nu}\dot{x}^\mu\dot{x}^\nu=-1$ leads to a first-order `effective potential' equation for $x$ upon eliminating $\dot{t}$, $\dot{z}$, and $\dot{w}$ using Eq.~\Eqref{particle_eom}:
\begin{align}
 &\dot{x}^2+U_{\mathrm{eff}}=0,\quad\mbox{where}\nonumber\\
 &U_{\mathrm{eff}}=-\brac{\sqrt{1-\sigma^2x^2}\mathcal{E}-e\sigma x}^2+1+\frac{\expo{-2\arcsin\sigma x}P_z^2+\expo{2\arcsin\sigma x}P_w^2}{1-\sigma^2x^2}.\label{particle_Ueff}
\end{align}
As mentioned earlier, charged particle motion in the magnetic V\v{Z} case has been studied in \cite{Vesely:2022vws}. Vesel\'{y}'s analysis should therefore agree with the present electric case only for neutral particles. In particular, one can check that for $e=0$, Eq.~\Eqref{particle_Ueff} agrees with Eq.~(2.186) of \cite{Vesely:2022vws}, where the latter uses coordinates $\rho$ where $1-\sigma^2x^2=\cos^2(\sigma\rho)$.

Another equation of motion for $x$ can be obtained from the Euler--Lagrange equation, $\frac{\dif}{\dif\tau}\frac{\partial L}{\partial\dot{x}}=\frac{\partial L}{\partial x}$, which leads to the second-order differential equation
\begin{align}
 \ddot{x}&=-\sigma^2x\brac{\sqrt{1-\sigma^2x^2}\mathcal{E}-e\sigma x}^2-\frac{e\sigma}{\brac{1-\sigma^2x^2}}\brac{\sqrt{1-\sigma^2x^2}\mathcal{E}-e\sigma x}\nonumber\\
  &\hspace{1cm}+\frac{\sigma\brac{\sqrt{1-\sigma^2x^2}-\sigma x}\expo{-2\arcsin\sigma x}}{\brac{1-\sigma^2x^2}^2}P_z^2-\frac{\sigma\brac{\sqrt{1-\sigma^2x^2}+\sigma x}\expo{2\arcsin\sigma x}}{\brac{1-\sigma^2x^2}^2}P_w^2.\label{xddot}
\end{align}
Equation \Eqref{xddot} can be integrated using the Runge--Kutta methods and used as a numerical check to verify that Eq.~\Eqref{particle_Ueff} is always satisfied throughout the motion.

From Eqs.~\Eqref{zdot} and \Eqref{wdot}, we see that $\dot{z}$ and $\dot{w}$ does not change sign throughout the motion. Therefore a particle with non-zero momenta $P_z$ and $P_w$ will simply evolve monotonically in the $z$ and $w$ directions. Therefore in the following, we will be mainly interested in the $x$-motion, and shall set $P_z=P_w=0$. 

Note that since $\dot{x}^2+U_{\mathrm{eff}}=0$, the particle is confined to the domain where $U_{\mathrm{eff}}\leq 0$. Assuming $q\geq 0$, this condition requires that particles are confined to the region (for $P_z=P_w=0$)
\begin{align}
 x_-\leq x\leq x_+,\quad\mbox{where}\quad x_\pm=\frac{-\brac{e\mp\mathcal{E}\sqrt{\mathcal{E}^2+e^2-1}}}{\brac{\mathcal{E}^2+e^2}\sigma}.
\end{align}
If either $|x_+|\geq a$ or $|x_-|\geq a$, the particles have sufficient kinetic energy to collide with either of the capacitor plates. Conversely, if $|x_\pm|<a$, the particle will oscillate in a finite region in between the plates. 

The effective potential $U_{\mathrm{eff}}$ has a local minima (for $P_z=P_w=0$) at 
\begin{align}
 x_{\mathrm{eq}}=\frac{q}{\sigma\sqrt{\mathcal{E}^2+q^2}},
\end{align}
so that particles with energy and charges obeying $\left.U_{\mathrm{eff}}\right|_{x=x_{\mathrm{eq}}}=0$ remain in a stable equilibrium. In particular for neutral particles the minima of the potential is at $x_{\mathrm{eq}}=0$, reflecting the fact that the V\v{Z} spacetime is a symmetric gravitational potential well.

\begin{figure}
 \centering 
 \includegraphics{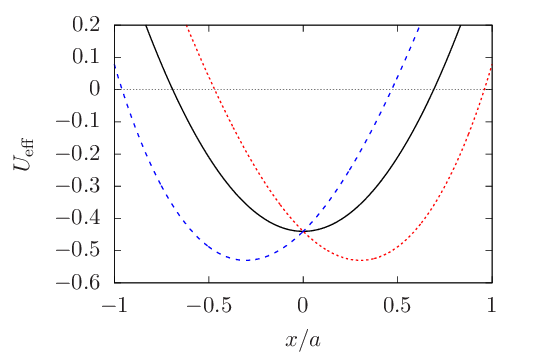}
 \caption{Plot of $x$ vs $U_{\mathrm{eff}}$ for $\mathcal{E}=1.2$, $P_z=P_w=0$, and $\sigma=0.8/a$ for charges $e=0$ (solid black curve), $e=0.3$ (dashed blue curve) and $e=-0.3$ (dotted red curve).}
 \label{fig_Ueff}
\end{figure}

In Fig.~\ref{fig_Ueff}, the plots of $x$ vs $U_{\mathrm{eff}}$ are shown for the case $\mathcal{E}=1.2$ and $\sigma=0.8/a$ for various charges.  We see that for neutral particles ($e=0$, the solid black curve in Fig.~\ref{fig_Ueff}.) the effective potential has a minimum at $x=0$. As mentioned above, this shows that the V\v{Z} spacetime is a symmetric gravitational potential well. For positive charges ($e=0.3$ in Fig.~\ref{fig_Ueff}, dashed blue curve.) the minimum of $U_{\mathrm{eff}}$ is shifted towards the left, reflecting the attraction of the positive charge towards the negative plate, as well as the repulsion from the positive one. Similarly, negative charges ($e=-0.3$ in Fig.~\ref{fig_Ueff}, dotted red curve.) are attracted towards the positive plate and repulsed by the negative one, shifting the equilibrium point towards the left. 

\section{Spherical capacitor} \label{sec_spherical}

In Sec.~\ref{sec_eom}, it was found that the parallel-plate capacitor requires the support of an external matter configuration in order to satisfy the null energy condition (NEC). The energy density $\mathcal{T}_{tt}$ is negative, so the weak energy condition (WEC) is still not satisfied. Furthermore, the simplest case of the V\v{Z} spacetime sandwich between two flat Minkowski spacetimes will violate even the NEC, hence requiring exotic matter to realise this configuration.

In this section, we consider a configuration with spherical symmetry, where the spacetime carrying the electric field is that of Reissner--Nordstr\"{o}m. Shell sources for the Reissner--Nordstr\"{o}m spacetime have been studied in detail in \cite{Lemos:2011dq,Uchikata:2012zs,Fernandes:2022gjd}. Here, we will briefly show that it is possible to obtain a capacitor configuration that satisfies the WEC.

\begin{figure}
 \centering
 \includegraphics{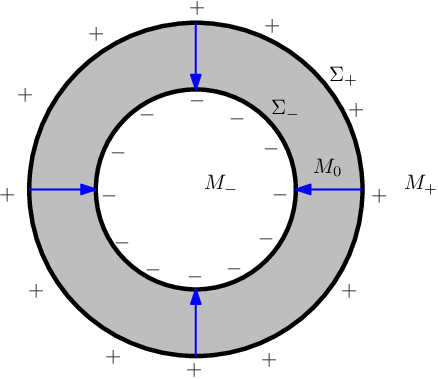}
 \caption{Sketch of a spherical capacitor. }
 \label{fig_SphericalC}
\end{figure}

As before, let us write spacetime as three connected pieces $M=M_-\cup M_0\cup M_+$, but now with spherical symmetry, as sketched in Fig.~\ref{fig_SphericalC}. The bulk solution for $M_0$ will be the Reissner--Nordstr\"{o}m spacetime, where the metric and potential is
\begin{subequations}
\begin{align}
 \dif s^2_0&=-f(r)\dif t^2+f(r)^{-1}\dif r^2+r^2\dif\Omega^2_{(2)},\quad f(r)=1-\frac{2\mu}{r}+\frac{Q^2}{r^2},\\
  A&=\frac{Q}{r}\,\dif t,
\end{align}
\end{subequations}
where $\dif\Omega^2_{(2)}$ is the metric of a round 2-sphere. We'll take the domain for the radial coordinate to be $r_-\leq r\leq r_+$, and we assume no roots of $f$ occur in this domain. That is, we assume no horizons in $M_0$. For the exteriors, we shall take $M_\pm$ to be Schwarzschild spacetimes,
\begin{subequations}
\begin{align}
 \dif s_\pm&=-V_\pm(r)\alpha_\pm\dif t^2+V_\pm(r)^{-1}\dif r^2+r^2\dif\Omega_{(2)}^2,\quad V_\pm(r)=1-\frac{2m_\pm}{r},\\
   A_\pm&=\pm\frac{Q}{r_\pm}\,\dif t,
\end{align}
\end{subequations}
where $\alpha_\pm=f(r_\pm)/V_\pm(r_\pm)$ is a constant fixed by the first junction condition. As before, $A_\pm$ is constant on $M_\pm$ and hence there are no electric fields.

Applying the methods outlined in Sec.~\ref{sec_eom} to this case, we find that the surface stress tensors on $\Sigma_\pm$ are respectively
\begin{align}
 \mathcal{T}_{tt}^\pm=\rho^\pm f,\quad \mathcal{T}_{ij}^\pm=p^\pm r^2\tilde{\gamma}_{ij},
\end{align}
where
\begin{subequations}
\begin{align}
 \rho^\pm&=\pm\frac{1}{8\pi}\frac{2}{r_\pm}\left.\brac{-V_\pm^{1/2}+f^{1/2}}\right|_{r_\pm},\\
 p^\pm&=\pm=\frac{1}{8\pi}\left.\sbrac{\frac{1}{r_\pm}\brac{V^{1/2}_\pm-f^{1/2}}+\half\brac{\frac{V_\pm'}{V_\pm^{1/2}}-\frac{f'}{f^{1/2}}}}\right|_{r_\pm},
\end{align}
\end{subequations}
where the functions $V_\pm$, $f$, and their derivatives are evaluated at $r=r_\pm$. In this case, the WEC can be satisfied by appropriate choices of $m_\pm$. Furthermore, using the invariant definitions for charge and potential difference, the capacitance is
\begin{align}
 C=4\pi\epsilon_0\brac{\frac{1}{r_+f(r_+)^{1/2}}-\frac{1}{r_-f(r_-)^{1/2}}}^{-1}.
\end{align}
Unlike the parallel-plate case, the spherical capacitance receives modification due to gravity. In the weak-gravity limit, $f(r_\pm)\rightarrow 1$ and we recover the elementary formula for spherical capacitance.

\section{Conclusion} \label{sec_conclusion}

In this paper we have removed the curvature singularity of the V\v{Z} spacetime and sandwiched it with asymptotically Minkowski spacetimes on both sides. This configuration is interpreted as a parallel-plate capacitor, where junction conditions determine the properties of the capacitor plates. 

Defining the capacitance (per area) $\mathcal{C}$ as the ratio of the charge (per area) with the potential difference between the two plates, we find, in SI units, the expression $\mathcal{C}=\epsilon_0/2a$ for the parallel-plate capacitor, which is the same as the elementary non-relativistic formula. However, for the spherical case, the capacitance is $C=4\pi\epsilon_0\brac{\frac{1}{r_+f(r_+)^{1/2}}-\frac{1}{r_-f(r_-)^{1/2}}}^{-1}$, where $f(r)$ is the Reissner--Nordstr\"{o}m lapse function and is less than $1$ for strong gravity. In other words, unlike the parallel-plate capacitor, the spherical capactance is modified when spacetime curvature is taken into account.

%The factor $(1-\sigma^2a^2)^{-1}$ represents the gravitational contribution to the capacitance, as a non-zero $\sigma x\sim \sigma a$ signifies spacetime curvature in the metric. Therefore, we find that gravitational effects behaves as an effective dielectric constant which modifies the capacitance by $\epsilon_0\rightarrow\epsilon_0(1-\sigma^2a^2)^{-1}$. In the weak gravity limit, we have $\sigma a\rightarrow0$ and the elementary formula for the capacitance is recovered.

Perhaps an undesirable feature of the parallel-plate capacitor is that the junction conditions requires the surface stress tensor to have negative energy density, leading to a violation of the weak energy condition. However, by an appropriate choice of the exterior matter distribution, it is possible to find a configuration that obeys the null energy condition. On the other hand, the spherical capacitor sandwiched between two Schwarzschild spacetimes does satisfy the weak energy condition.

The presence of gravity between the plates contributes to some non-trivial behaviour for charged test particles between the parallel-plate capacitor. In a non-gravitational capacitor, any charged particle would simply be uniformly accelerated by the electric field towards the plate of opposite charge. However, in the V\v{Z} region, there exist a potential well due to the gravitational force. So in this case, it is possible for charged particles to simply oscillate in between the plates without touching them.

\section*{Acknowledgments}

The authors thank Tomasz Paterek and Kalai Kumar a/l Rajagopal for illuminating discussions. Y.-K.~L is supported by Xiamen University Malaysia Research Fund (Grant no. XMUMRF/2021-C8/IPHY/0001).

\bibliographystyle{CosmicC}

\bibliography{CosmicC}

\end{document}